%% file: frustration-v20.tex
\documentclass[prx,aps,reprint,floatfix]{revtex4-2} 
\usepackage{graphicx} \usepackage{subfigure}
\usepackage{stdclsdv} \usepackage{array,booktabs} \usepackage{amsmath}
\usepackage{amsfonts}
\usepackage{tabularx}
\usepackage[english]{babel}
\usepackage{microtype}

\newcommand{\Fig}[1]{Figure~\ref{#1}}
\newcommand{\Tab}[1]{Table~\ref{#1}}

\newcommand{\Sec}[1]{Section~\ref{#1}}

\newcommand{\Eq}[1]{Equation~\ref{#1}}
\newcommand{\Eqs}[1]{Equations~\ref{#1}}

\newcommand{\stra}{s}
\newcommand{\strb}{{s'}}

\begin{document}

\title{The Range of Geometrical Frustration in Lattice Spin Models}

\author{Pierre Ronceray}
\affiliation{Princeton Center for Theoretical Science, Princeton University, Princeton, NJ 08544, USA}
\author{Bruno \surname{Le Floch}}
\affiliation{Philippe Meyer Institute, Physics Department, \'Ecole Normale Sup\'erieure, PSL Research University, Paris, France}

\begin{abstract}
  The concept of geometrical frustration in condensed matter physics
  refers to the fact that a system has a locally preferred structure
  with an energy density lower than the infinite ground state. This
  notion is however often used in a qualitative sense only. In this
  article, we discuss a quantitative definition of geometrical
  frustration in the context of lattice models of binary spins. To
  this aim, we introduce the framework of local energy landscapes,
  within which frustration can be quantified as the discrepancy
  between the energy of locally preferred structures and the ground
  state. Our definition is scale-dependent and involves an
  optimization over a gauge class of equivalent local energy
  landscapes, related to one another by local energy
  displacements. This ensures that frustration depends only on the
  physical Hamiltonian and its range, and not on unphysical choices in
  how it is written. Our framework shows that a number of popular
  frustrated models, including the antiferromagnetic Ising model on a
  triangular lattice, only have finite-range frustration: geometrical
  incompatibilities are local and can be eliminated by an exact
  coarse-graining of the local energies.
\end{abstract}

\maketitle

Frustration refers to the situation in which the simultaneous
minimization of all local interaction energies in a system is not
possible, due to the incompatibility of local
constraints~\cite{toulouse_theory_1977}. We can distinguish here the
cases in which this frustration is forced by the imposition of a
frozen disorder in the form of random fields or interactions (such as
in spin glasses~\cite{anderson_concept_1978}) from those in which the
frustration arises directly from an intrinsic mismatch in the uniform
interactions between constituents. The latter situation, referred to
as geometrical
frustration~\cite{tarjus_frustration-based_2005,sadoc_geometrical_2006,charbonneau_geometrical_2013},
is the topic of this article. This definition is essentially
conceptual and qualitative, although some system-specific quantitative
measurements exist, such as the measure of a spontaneous curvature of
hard sphere
systems~\cite{sadoc_geometrical_2006,tarjus_frustration-based_2005} or
the incompatibility between spontaneous splay and bend in bent-core
liquid crystals~\cite{niv_geometric_2018}.  It is often rephrased in
the following way: the locally preferred structure, which results from
local minimization of the energy, cannot tile the whole space. Note
that we consider here constraints intrinsic to the geometry of the
local order, but not surface effects induced by a mismatch at the
boundaries of the system.

In this article, we examine this notion of geometrical frustration --
\emph{i.e.\@} the incompatibility of the best local order with
space-filling -- and attempt at making it quantitative within the
realm of lattice spin models (without quenched disorder). We start in
\Sec{sec:examples} by motivating this work through the study of
frustration in two simple lattice models, which reveal two caveats for
a quantitative measure of frustration: \emph{(i)} it depends on the
scale considered, and \emph{(ii)} it should not be affected by
\emph{energy displacements}, a type of gauge transformation that
locally redistribute the energy while leaving the total Hamiltonian
unaffected. In \Sec{sec:formal}, we then address these challenges and
propose a formalism, \emph{Local Energy Landscapes}, within which, we
argue, geometrical frustration can be well-defined. This allows us to
distinguish two classes of frustrated systems: in most models,
including the archetypal antiferromagnetic Ising model on a triangular
lattice, frustration has a \emph{finite range}
and can be eliminated in a single exact
coarse-graining step. In other cases, it could persist at all scales,
a behavior we term \emph{long-range frustration}.  Our framework allows
to distinguish these qualitatively distinct facets of frustration, and
quantitatively measure it in a way that depends only on the scale
considered and on the global Hamiltonian, not on unphysical details.

\section{Two case studies}
\label{sec:examples}

To motivate our study, and in particular illustrate the difficulties
encountered when attempting to define a quantitative measure for
frustration, we first discuss frustration in two simple models.

\subsection{The antiferromagnetic Ising model}
\label{sec:AFI}

\begin{figure}[bt]
  \centering
  \includegraphics[width=\columnwidth]{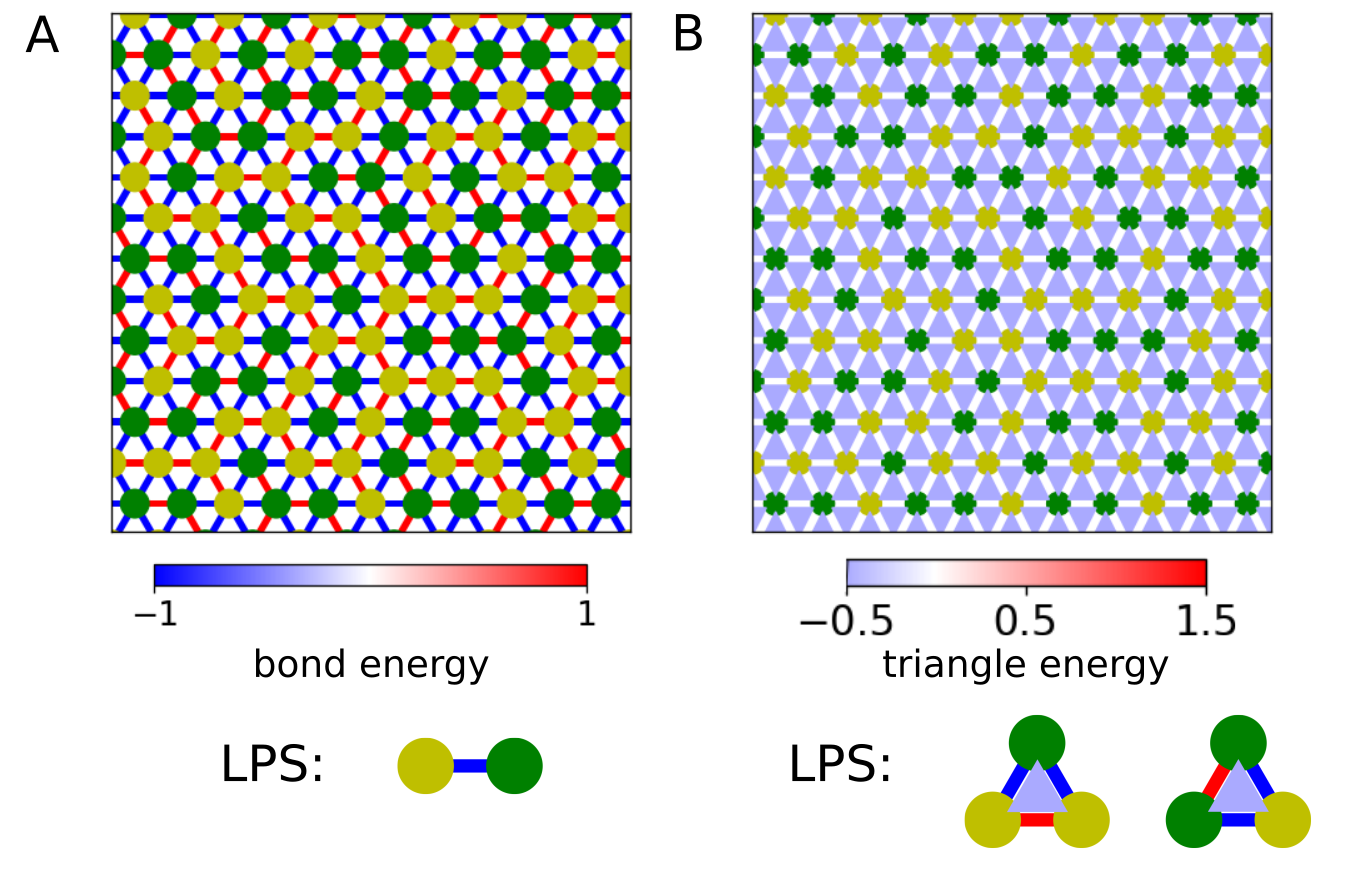}
  \caption{\textbf{A}. A ground state configuration of the
    antiferromagnetic Ising model on a triangular lattice
    (\Eq{eq:AFI_bonds}). The color of sites indicates their state of
    spin. The color of bonds indicates their energy; blue bonds
    correspond to the locally preferred structure (LPS). Red bonds are
    defects due to frustration. \textbf{B.} The same configuration,
    now showing the energy of triangles as in
    \Eq{eq:AFI_triangles}. All triangular plaquettes are in the LPS,
    and the model appears to be unfrustrated. }
  \label{fig:AFI}
\end{figure}

We start by examining what is probably the most popular example of
frustrated system~\cite{diep_frustrated_2013}: the antiferromagnetic
Ising model on a triangular lattice (\Fig{fig:AFI}). Its Hamiltonian
reads
\begin{equation}
  \label{eq:AFI_bonds}
  H = \sum_{i\sim j} s_i s_j
\end{equation}
where the sum runs over all edges of the lattice, and $s_i = \pm 1$
are the local spin variables. The ground state energy per site of this
model is $E_0 = -1$ (\Fig{fig:AFI}A). However, minimizing
independently each term in the sum of \Eq{eq:AFI_bonds} would result
in an energy per site of $E^*_\mathrm{bonds} = -3$, corresponding to
each edge having an energy of $-1$. The ``locally preferred order'',
corresponding to antiparallel spins, is thus frustrated, as
$E^*_\mathrm{bonds}<E_0$: it cannot be simultaneously achieved at all
edges, due to the presence of triangles that overconstrain the
system~\cite{toulouse_frustration_1980,diep_frustrated_2013}. A simple
quantification of frustration in this model would thus be
$f_\mathrm{bonds} = E_0-E^*_\mathrm{bonds} = 2$, \emph{i.e.\@}  the
difference between the energy per site in the ground state, and that
in an ideal state where the preferred local order would be achieved
everywhere. This frustration is generally invoked as the cause of the
extensive degeneracy of the ground state of this system~\footnote{Note
  that an extensively degenerate ground state --\emph{i.e.\@} a non-zero
  entropy at zero temperature -- is sometimes considered to be the
  \emph{definition} of frustration, rather than one of its effects. We
  will not take that point of view here. }.

This definition is not without danger, however: indeed, consider the
following rewriting of the Hamiltonian,
\begin{equation}
  \label{eq:AFI_triangles}
  H = \sum_{\mathrm{triangles}\ (ijk)} \tau_{ijk}
\end{equation}
where the sum runs over triangles of three bonds, and we define
$\tau_{ijk} = \frac{1}{2} (s_is_j+s_js_k + s_ks_i)$ as the energy of
such a triangle. As each bond is part of two triangles,
\Eqs{eq:AFI_bonds} and \ref{eq:AFI_triangles} are clearly two
equivalent ways of writing the same Hamiltonian. However, minimizing
each term independently in \Eq{eq:AFI_triangles} now results in an
energy per site of $E^*_\mathrm{triangles} = -1$, corresponding to
each triangle having the minimum possible energy of $t_{ijk} = -1/2$
(\Fig{fig:AFI}B). We thus have
$f_\mathrm{triangles} = E_0-E^*_\mathrm{triangles} = 0$: the
Hamiltonian written in \Eq{eq:AFI_triangles} is unfrustrated, as its
locally preferred order can tile the whole lattice. From this point of
view, this system is extensively degenerate because it is
underconstrained: as in some plaquette models, the simultaneous
minimization of all terms of the Hamiltonian is not sufficiently
constraining to select a single periodic ground
state~\cite{diep_frustrated_2013}.

These two ways of writing the same Hamiltonian thus lead to different
conclusions as to whether it is frustrated or not. Clearly, there is
more information in Hamiltonian~\ref{eq:AFI_bonds} in terms of the
locality of the energy: \Eq{eq:AFI_triangles} is a less local way of
writing the energy, and its energy density can be seen as an exact
coarse-graining of the energy density of \Eq{eq:AFI_bonds}, by
averaging the energy of each triangle. Since this coarse-graining
removes frustration, we can qualify this type of frustration of
\emph{finite range},
or irrelevant: it vanishes under renormalization. In order to
quantify frustration in this system, one should therefore specify what
scale is being considered: the antiferromagnetic Ising model on the
triangular lattice is frustrated when going from the scale of a single
bond to a triangle, but not from the scale of a triangle to the
infinite lattice.

\subsection{A minimal frustrated model?}
\label{sec:FNM}

\begin{figure}[bt]
  \centering
  \includegraphics[width=0.9\columnwidth]{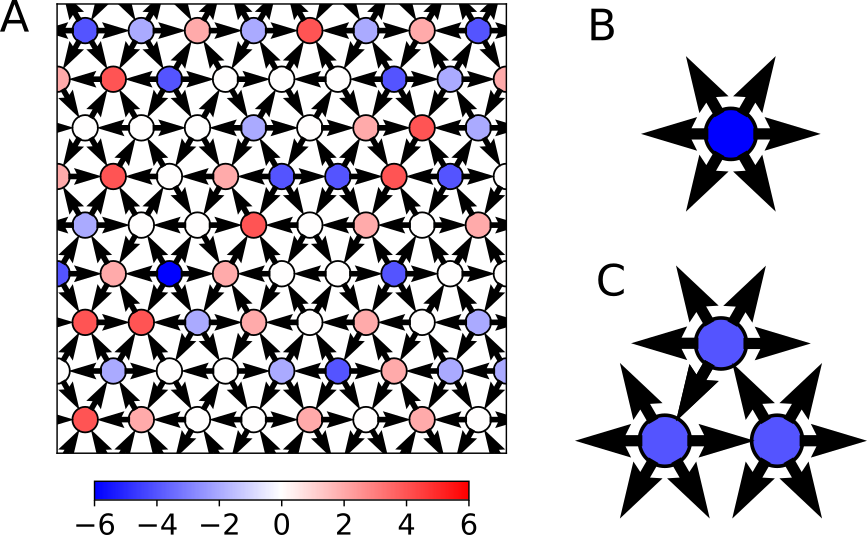}
  \caption{The ``frustrated model'' defined by \Eq{eq:FNM}. The
    degrees of freedom are the orientation of the edges connecting two
    nodes. \textbf{A.} A typical ground state configuration. The color
    of each node indicates its energy $\phi_i$. \textbf{B.} The
    locally preferred structure at the scale of a single node has an
    energy $-6$. \textbf{C.} At the scale of three nodes, the LPS has
    energy $-4$ per site. }
  \label{fig:FNM}
\end{figure}

We now discuss a second simplistic model that exhibits, we suggest,
surprising frustration properties. Consider a triangular lattice where
each bond carries a binary variable of orientation -- pointing towards
either of the two sites it connects (\Fig{fig:FNM}A). We define the
following Hamiltonian for this system:
\begin{equation}
  \label{eq:FNM}
  H_\mathrm{FNM} = \sum_i \phi_i
\end{equation}
where the local energy $\phi_i$ is the difference between the number
of edges attached to $i$, pointing towards $i$, to the number of edges
pointing against $i$ -- \emph{i.e.\@} the local flux at $i$. This is a
specific instance of the $64$-vertex
model~\cite{diep_frustrated_2013}. The locally preferred structure
corresponds to six edges pointing away from $i$ (\Fig{fig:FNM}B), and
tiling the lattice with such sites would result in an energy of
$E^*_\mathrm{site} = -6$. This is however impossible, and ground state
configurations include many defects to this ideal structure
(\Fig{fig:FNM}A): this system is frustrated.

Grouping local energy variables together, as we did in
\Eq{eq:AFI_triangles}, reduces the frustration but does not cancel it:
the locally preferred order at the scale of a triangle of three sites
has energy per site $E^*_\mathrm{triangles} = -4$, still higher than
the ground state (\Fig{fig:FNM}C). This can be easily generalized to any
cluster of sites: frustration in this model thus appears to be
long-range, that is, it cannot be blurred out by a
coarse-graining. This model has many peculiar properties, such as
extensive degeneracy of the ground state, characterized for instance
by the fact that the reversal of any closed loop of edge variables
leaves the energy unchanged.

Rather than leading the reader further on, let us examine more closely
the Hamiltonian proposed in \Eq{eq:FNM}. Each edge contributes to two
$\phi_i$ variables, each with an opposite sign: reversing its
orientation thus displaces energy from one site to the other, but
leaves the total energy unchanged -- specifically, each edge variable
has a zero contribution to the total energy, and thus \Eq{eq:FNM} can
be rewritten exactly as
\begin{equation}
H_\mathrm{FNM} = 0.\label{eq:FNM0}
\end{equation}
This model thus has the appearance of being frustrated, while being
completely trivial -- in a sense, it is a \emph{Frustrated Non-Model}
(FNM)... Admittedly, the Hamiltonian in \Eq{eq:FNM} is quite simple,
and an aware reader could have realized that its frustration is only
superficial.  However, for an observer who only has access to the
$\phi_i$ variables and the resulting field of local energies (\Fig{fig:FNM}A),
this is far from being obvious.

The field of local energies $\phi_i$ as defined here consists in what
we define as an \emph{energy displacement}, \emph{i.e.\@} a
configuration-dependent spatial patterning of the energy that always
has zero sum, and thus no influence on the total
Hamiltonian. Importantly, adding such an energy displacement to any
non-trivial Hamiltonian would leave it unchanged: it would change
``local energies'', but not the total energy of any state -- and hence
result in identical dynamical and thermodynamical properties. Two
models that differ by a local energy displacement are thus physically
and mathematically equivalent: their only difference lies in the way
that the Hamiltonian is written in terms of local energies -- a
distinction that is arguably unphysical, and can be compared to a
gauge change.  However, as local energies are modified by energy
displacements, they can affect the energy and even the nature of the
locally preferred structure. This significantly complicates the
problem of quantifying frustration. Indeed, any useful and physically
meaningful definition of frustration should be gauge invariant and
depend only on the Hamiltonian, not on the specific way that it is
written -- it should, in particular, see through \Eq{eq:FNM} and
consider this model as non-frustrated, as its equivalent formulation
in \Eq{eq:FNM0} is trivial. Note that there is a scale to such energy
displacements: as they change the local energies, they can also
effectively change the range of interactions of the
Hamiltonian. Indeed, in the case of the antiferromagnetic Ising model,
going from \Eq{eq:AFI_bonds} to~\ref{eq:AFI_triangles} can be seen as
an energy displacement, moving the energy from the bonds to the
triangles.

\section{Frustration of Local Energy Landscapes}
\label{sec:formal}

In the previous section, we have identified two caveats that should be
addressed in order to quantify geometrical frustration in a meaningful
way. First, frustration should be a function of scale: as the
structures considered get larger, the locally preferred structure will
resemble more and more the ground state, as it internalizes
constraints. Second, at a given scale, frustration should not depend
on the specific way that the Hamiltonian is written -- \emph{i.e.\@} it
should not be affected by a gauge change of the field of local
energies, corresponding to local energy displacements. We now propose
a framework to define and measure scale-dependent frustration. This
framework relies on the classification of all possible local
structures of the model at a given scale, and considering the ways to
attribute an energy to each of them -- \emph{i.e.\@} the ways to define
the \emph{Local Energy Landscape} (LEL).

This approach is common in the study of supercooled liquids and glassy
systems, where the idea of studying local structures is that a finite
number of geometries can accurately describe the local environments of
particles in a liquid: in supercooled liquids, distortions around
local energy minima can be neglected in first approximation, and it
makes sense to consider the energy of these local
structures~\cite{frank_supercooling_1952,tarjus_frustration-based_2005,royall_role_2015,taffs_role_2016}. This
point of view is even more relevant in lattice models of discrete
spins, in which the local structures are truly discrete: in such
cases, models with short-range interactions can be exactly expressed
in terms of their LEL, \emph{i.e.\@} the energy
associated to each possible local
structures~\cite{ronceray_liquid_2016}.

In this section, we first define the set of local structures
corresponding to a given scale (\Sec{sec:LS}), and introduce the Local
Energy Landscapes framework that maps structures onto local energies
(\Sec{sec:LEL}). We then show how to characterize the gauge of energy
displacements that change the LEL, but not the total Hamiltonian
(\Sec{sec:ED}). This allows us to propose a quantitative,
gauge-invariant measure of frustration at a given scale
(\Sec{sec:frustration}). We then discuss the scale-dependence of this
measure of frustration (\Sec{sec:range}). Finally, we discuss a
practical application of this method in the identification of ground
state energies of spin systems (\Sec{sec:GS}). In each subsection, we
first discuss concepts in their generality, then apply them to the
practical case of triangular lattices.

\input{clusters_table}

\begin{figure*}[bt]
  \centering
  \includegraphics[width=0.8\textwidth]{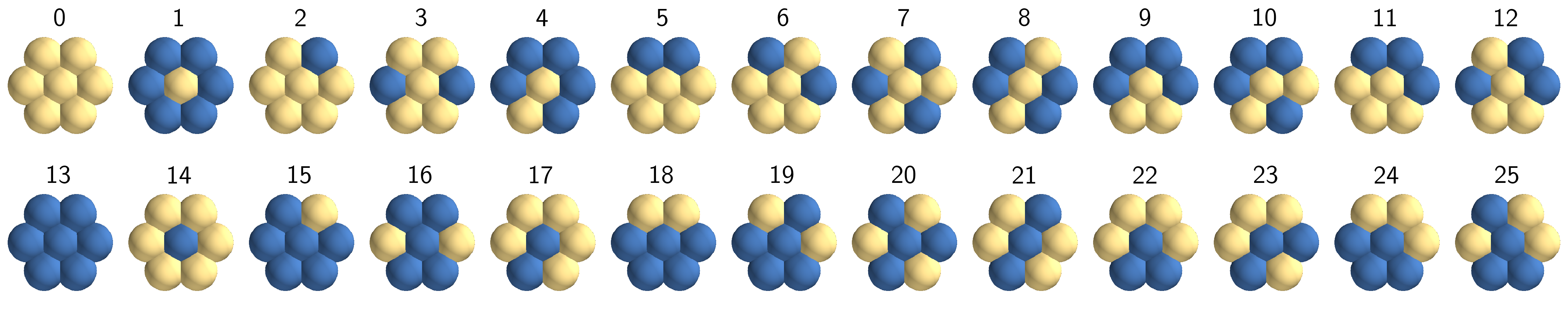}
  \caption{The $n=26$ distinct local structures corresponding to the
    coordination cluster with $z=7$ sites. The second row is
    spin-reversed compared to the first. Note that structures $12$ and
    $25$ are chiral; we treat both enantiomers as the same structure
    here. Distinguishing enantiomers would lead to two more local
    structures.}
  \label{fig:LSs}
\end{figure*}

\subsection{Local structures}
\label{sec:LS}

We consider a system of $N$ binary spins on a Bravais lattice with
periodic boundary conditions, in the limit of a large number of sites
$N\to\infty$ where the effect of boundaries becomes irrelevant, and
translation-invariant Hamiltonians with finite-range interactions on
this spin system.  All sites are thus equivalent, and can be
characterized by their spin environment, with which they interact. To
classify these environments, we first decide on a cluster of sites on
which we will define local structures. This cluster should be larger
than the interaction range of the Hamiltonian; the size $z$ of this
cluster (the number of sites it contains) sets the scale at which we
define and study frustration. It is convenient to use a cluster that
has the highest possible symmetry, as this will limit the number of
structures to consider. In \Tab{tab:clusters} we present a selection
of high-symmetry clusters on the triangular lattice. In most of this
article, we will take the coordination shell cluster (one site and its
six neighbors, $z=7$) as example to illustrate the concepts we
discuss.

Having chosen a cluster of sites, we now introduce the set of all
possible local structures on this cluster, \emph{i.e.\@} the possible
spin patterns on this cluster. There are $2^z$ distinct patterns,
however if the Hamiltonian is isotropic (\emph{i.e.\@} invariant under
the discrete lattice rotations) it makes sense to consider two
structures that differ by a rotation as identical. Depending on
whether the considered Hamiltonian is chiral, one can choose to treat
enantiomeric structures (\emph{i.e.\@} non identical mirror copies) as
distinct structures or not. Using these symmetries results in a set of
$n$ distinct local structures. The values of $n$ corresponding to each
cluster are presented in \Tab{tab:clusters}. In the case of the
triangular coordination shell, the $n=26$ structures are depicted in
\Fig{fig:LSs}.

A spin configuration of the system can be described by its
\emph{structural composition} $\mathbf{c}$, an $n$-dimensional vector
that specifies the fraction of sites in each local structure. The
number of sites in structure $\stra$ is thus $N_\stra = Nc_\stra$ in this
configuration. Note that a configuration is not fully characterized by
its structural composition; conversely, as we will see, not all
compositions are possible. Nevertheless, since the range of the
Hamiltonian is shorter or equal than the size of the cluster we
consider, the energy of a configuration is completely determined by
the corresponding structural composition vector $\mathbf{c}$.

\subsection{Local energy landscapes}
\label{sec:LEL}

We now introduce the \emph{local energy landscape} (LEL)
that relates the structural composition to the energy of the system.
We associate an energy $\epsilon_\stra$ to each site in structure $\stra$,
such the energy per site of the system reads
\begin{equation}
  \label{eq:H}
  E(\mathbf{c}) = \sum_{\stra=1}^n c_\stra \epsilon_\stra = \mathbf{c} \cdot \epsilon
\end{equation}
where the vector $\epsilon = \{ \epsilon_\stra\}_{\stra=1..n}$ is
the LEL of the system. A vast class of popular models can be written
exactly in such a form, which includes the Ising model, its variants
with antiferromagnetic and/or next-to-nearest neighbor interactions,
and plaquette models. The LEL thus fully characterizes the energetics
of the system in terms of local structures. While structures are not
energetically coupled in \Eq{eq:H}, it is important to note that they
are not independent: each spin is part of several structures, and
these overlaps result in entropic coupling between structures.
Indeed, the system's free energy per site at temperature $T$ can be
written (setting $k_B = 1$)~\cite{ronceray_liquid_2016}:
\begin{equation}
  \label{eq:F}
  F(T) = \min_{\mathbf{c}, S(\mathbf{c}) \geq 0} [ \mathbf{c}\cdot\mathbf{\epsilon} - T S(\mathbf{c}) ]
\end{equation}
where $S(\mathbf{c})$ is the entropy per site of a system with
structural composition $\mathbf{c}$, which effectively counts the
states available in the model compatible with these fractions of local
structures. By convention we have $S(\mathbf{c})=-\infty$ if there are
no states compatible with the structural composition $\mathbf{c}$ --
for instance if it violates the basic constraints that all $c_\stra$'s are
non-negative and that $\sum_\stra c_\stra =1$. The appeal of this approach
lies in the fact that $S(\mathbf{c})$ depends only on the lattice
geometry and the choice of cluster, not on the LEL. The thermodynamics
of a broad class of models can thus be related, by Legendre
transformation (\Eq{eq:F}), to a single function $S(\mathbf{c})$.
From this point of view, finding the ground state energy
$E_0(\epsilon)$ of a model with LEL $\epsilon$ corresponds to finding
the extremal point of definition of $S(\mathbf{c})$ along the
direction $\epsilon$,
\begin{equation}
  E_0(\epsilon) = \inf_{\mathbf{c}, S(\mathbf{c}) \geq 0}  \mathbf{c}\cdot\mathbf{\epsilon} 
  \label{eq:E0}
\end{equation}
which corresponds to the zero-temperature equilibrium state.

The minimum of $\epsilon$ corresponds to the minimal possible energy
of a site, \emph{i.e.\@} its energy when in the so-called \emph{Locally
  Preferred Structure} (LPS). This energy is a lower bound to the
ground state energy of the system:
\begin{equation}
  E_0(\epsilon) \geq \min_{\stra=1\dots n}  \epsilon_\stra
  \label{eq:E0_bound}
\end{equation}
When there is equality, the system is unfrustrated at the scale of the
structure considered: it can be uniformly tiled by locally preferred
structures. 

\subsection{The gauge of energy displacements}
\label{sec:ED}

The framework of local energy landscapes is significantly complicated
by the fact that \Eq{eq:H} is not sufficient to define the LEL: two
distinct local energy landscapes $\epsilon$ and $\epsilon'$ can indeed
correspond to the same physical system. This is the case if the
difference between them, $\delta = \epsilon'-\epsilon$, is an
\emph{energy displacement}, \emph{i.e.\@} a non-zero LEL corresponding
to a vanishing Hamiltonian. In this section, we show how to
characterize the set $\Delta$ of possible energy displacements
corresponding to a choice of local structures.

A LEL $\delta$ is an energy displacement if the energy of any possible
configuration, as given by \Eq{eq:H}, is zero:
$\mathbf{c}\cdot\delta = 0$ for all structural compositions
$\mathbf{c}$ such that $S(\mathbf{c})\geq 0$. The set $\Delta$ of
energy displacements $\delta$ thus has a vector space structure (a
linear combination of energy displacements still corresponds to a zero
Hamiltonian), which is a subspace $\Delta \subset \mathbf{R}^n$. A
linear analysis of the entropy functional $S(\mathbf{c})$ around the
infinite-temperature limit is sufficient to fully characterize this
vector space. Indeed, one can write the following expansion for the
entropy as a function of structural
composition~\cite{ronceray_geometry_2012,ronceray_influence_2013,ronceray_liquid_2016}:
\begin{equation}
  \label{eq:S_expansion}
  S(\mathbf{c}) = S_\infty - \frac{1}{2}(\mathbf{c}-\mathbf{c}^\infty)\cdot \mathbf{C}^{-1} \cdot (\mathbf{c}-\mathbf{c}^\infty) + O[(\mathbf{c}-\mathbf{c}^\infty)^3]
\end{equation}
where $S_\infty = \ln 2$ is the infinite-temperature entropy per site
of a binary spin system. Here $\mathbf{c}^\infty$ is the structural
composition at infinite temperature: in a fully random spin state, the
proportion of sites in structure $\stra$ is $c^\infty_\stra = g_\stra / 2^z$ where
$g_\stra$ is the number of rotational variants of structure $\stra$ and $z$ is
the number of sites in a structure~\cite{ronceray_geometry_2012}. The
matrix $\mathbf{C}$ in \Eq{eq:S_expansion} relates to structural
fluctuations at infinite temperature, and can be written as a
covariance matrix,
\begin{equation}
  \label{eq:covariance}
  C_{\stra\strb} = \frac{1}{N} \mathrm{Cov}_{T=\infty}(N_\stra,N_\strb)
\end{equation}
for a large system of $N$ sites, where $N_\stra$ is the number of sites in
structure $\stra$ in a given configuration. This matrix can either be
obtained by simulations, or computed exactly by enumerating overlaps
of structures~\cite{ronceray_geometry_2012,ronceray_liquid_2016}.

\begin{figure}[bt]
  \centering
  \includegraphics[width=0.8\columnwidth]{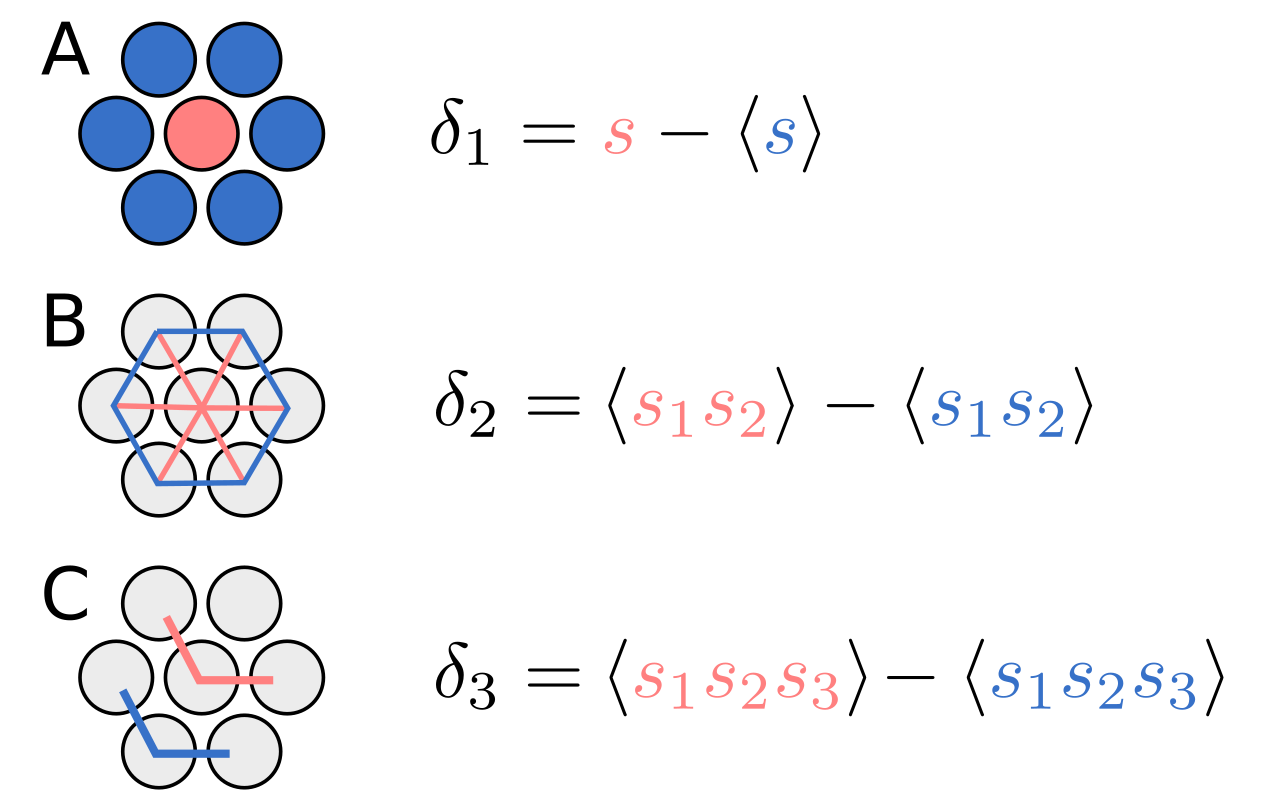}
  \caption{Schematic of the conserved quantities leading to the
    generators of the three-dimensional gauge space of energy
    displacements for the triangular coordination shell. \textbf{A.}
    Here $\delta_1$ is the difference between the spin value of the
    central site (red), to the average over its six neighbors
    (blue). This quantity is nonzero for all structures presented in
    \Fig{fig:LSs} except $0$ and $13$. However, the spatial average of the
    blue and the red term are both equal to the average spin value,
    for any configuration. Therefore, $\delta_1 \cdot \mathbf{c}=0$
    for all possible structural composition $\mathbf{c}$. \textbf{B.}
    Similarly, $\delta_2$ corresponds to the difference of two-spins
    interaction for radial edges (red) and lateral edges (blue) in the
    structure. While this can be locally nonzero, the spatial average
    of this quantity always vanishes. \textbf{C.} The last energy
    displacement corresponds to three-spin interactions that involve,
    or not, the central spin (averaged over all orientations). }
  \label{fig:ED}
\end{figure}

Importantly, the matrix $\mathbf{C}$ typically has a non-trivial null
space $\mathrm{Ker} (\mathbf{C})$, \emph{i.e.\@} the set of
eigenvectors associated to a zero eigenvalue. As \Eq{eq:S_expansion}
involves
$(\mathbf{c}-\mathbf{c}^\infty)\cdot \mathbf{C}^{-1} \cdot
(\mathbf{c}-\mathbf{c}^\infty)$, we have $S(\mathbf{c}) = -\infty$ for
any composition $\mathbf{c}$ for which $\mathbf{c}-\mathbf{c}^\infty$
has a nonzero projection on $\mathrm{Ker} (\mathbf{C})$: the expansion
of $S(\mathbf{c})$ ``detects'' forbidden configurations. The elements
of $\mathrm{Ker} (\mathbf{C})$ correspond to the existence of
conserved quantities. For instance, the constraint that
$\sum_\stra c_\stra =1$ (\emph{i.e.\@} that the set of structures
is complete) implies that $(1,1,\dots,1)\in\mathrm{Ker} (\mathbf{C})$
for all choices of local cluster. In the case of the triangular
coordination shell, there are also three non-trivial conservation
laws, corresponding to redundancies in one-, two- and three-body
interaction terms within the shell (see \Fig{fig:ED} and caption). As
a result, $\mathrm{Ker} (\mathbf{C})$ is four-dimensional for this
choice of local cluster.

The space $\Delta$ of energy displacements corresponds to vectors that
have $\delta \cdot \mathbf{c} = 0$ for all ``acceptable compositions''
$\mathbf{c}$ such that $S(\mathbf{c})\geq 0$. As we have seen, these
compositions are such that for any
$\lambda \in \mathrm{Ker} (\mathbf{C})$, we have
$\lambda \cdot (\mathbf{c}-\mathbf{c}^\infty) = 0$. The set of energy
displacements thus corresponds to vectors $\delta$ that are both
orthogonal to $\mathbf{c}^\infty$ and to all
$(\mathbf{c}-\mathbf{c}^\infty)$ for acceptable compositions
$\mathbf{c}$. Mathematically, we thus have:
\begin{equation}
  \label{eq:Delta}
  \Delta = \mathrm{Ker} (\mathbf{C})  \cap \mathrm{Perp} (\mathbf{c}^\infty)
\end{equation}
with $\mathrm{Perp} (\mathbf{c}^\infty)$ the hyperplane orthogonal to
the vector $\mathbf{c}^\infty$, and $\mathrm{Ker} (\mathbf{C})$ the
null space of $\mathbf{C}$. The exact values of both the covariance
matrix $\mathbf{C}$ and the infinite-temperature composition
$\mathbf{c}^\infty$ are analytically accessible; \Eq{eq:Delta} thus
provides an operational way to classify energy displacements for a
given definition of local structures. In Table~\ref{tab:clusters}, we
indicate the dimensionality of the vector space $\Delta$ for all
choices of local cluster. All clusters except the smallest ones admit
energy displacements.

To summarize, the space $\Delta$ of energy displacements acts as a
gauge group for the definition of local energy landscapes: for
$\delta\in\Delta$, the landscapes $\epsilon$ and $\epsilon+\delta$ are
physically equivalent, as they correspond to the same Hamiltonian. 

\subsection{A gauge invariant definition for frustration}
\label{sec:frustration}

We now examine the influence of energy displacements on the
quantification of frustration. As discussed in \Sec{sec:AFI}, one
should first specify the scale $z$ at which we define it,
corresponding to the number of sites in the cluster used to define the
local energy landscape. The idea of frustration as the incompatibility
between the Locally Preferred Structure (LPS) and filling space can be
intuitively quantified by the difference between the energy of a site
in the LPS, \emph{i.e.\@} the local energy landscape minimum
$\min_\stra \epsilon_\stra$, and the average energy per site in the ground state
$E_0$. This quantity, however, is not gauge invariant: two energy
landscapes corresponding to the same Hamiltonian may have different
minima. In particular, arbitrarily large ``apparent frustration'' can
be produced by adding a large energy displacement to any LEL, as we
made evident in \Sec{sec:FNM}.

We thus argue that the pertinent way to quantify the frustration of a
Hamiltonian is to \emph{minimize} it over the gauge group
$\Delta$. This way, a system will be considered unfrustrated if it
admits a LEL representation $\epsilon^*$ such that
$\min_\stra \epsilon^*_\stra = E_0$. If it does not, then the system is
frustrated, and the smallest gap $f_z$ between the ground state energy
and the energy of the LPS quantifies frustration at scale $z$:
\begin{equation}
  \label{eq:frustration}
  f_z(\epsilon) = E_0(\epsilon) - \max_{\delta\in\Delta}\left[\min_{\stra=1\dots n}(\epsilon_\stra+\delta_\stra) \right]
\end{equation}
Here $f_z(\epsilon)$ is a non-negative quantity, and the optimization
over $\delta$ ensures that it is gauge invariant. Note that the
quantity $\min_\stra(\epsilon_\stra+\delta_\stra)$, corresponding to the LPS
energy of the LEL $\epsilon+\delta$, is a linear-by-parts, concave
function of $\delta$: the maximization in \Eq{eq:frustration} is thus
non-ambiguous (there are no local maxima). While this optimization
might not be tractable analytically, it can be efficiently performed
numerically with algorithms such as the Nelder-Mead
simplex~\footnote{Importantly, symmetries of the original LEL $\epsilon$
  (for instance a discrete rotation or spin-flip symmetry) cannot be
  spontaneously broken when optimizing this concave function, so it is
  enough to consider energy displacements that preserve all symmetries.
  This makes the problem numerically tractable.}.

In Equation~\ref{eq:frustration}, the LEL
$\epsilon^* = \epsilon + \delta$ that maximizes the LPS energy plays a
special role. It typically has at least two distinct, degenerate LPS:
indeed, as the LPS energy is maximal, it should be such that no
further energy displacement can increase the energy of a LPS without
also decreasing the energy of the other.

\begin{figure}[bt]
  \centering
  \includegraphics[width=\columnwidth]{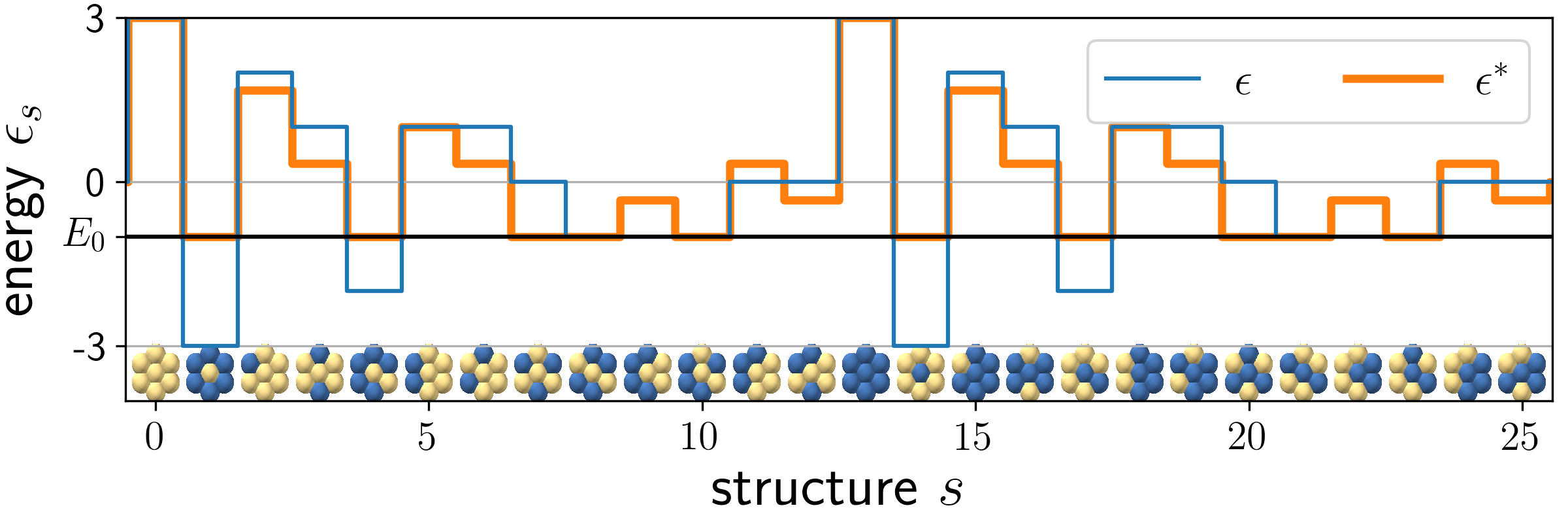}
  \caption{Two equivalent local energy landscapes, at the scale of the
    coordination shell, for the antiferromagnetic Ising
    model. \emph{Blue}: the local energy is
    $\frac{1}{2} s_i \sum_{j\sim i} s_j $, which corresponds to taking
    into account the energy of the radial bonds connecting $i$ to its
    neighbors. The local structures $1$ and $14$ are locally
    preferred, with energy $-3$, to compare with the ground state
    energy $E_0 = -1$. \emph{Orange}: the optimal LEL for this model,
    as defined by \Eq{eq:frustration}. The LPS is largely degenerate
    (1, 4, 7, 8, 10 and their spin-reversed variants), and all have
    energy $-1$, equal to the ground state energy. Any configuration
    composed exclusively of these structures is a ground state
    configuration; as many such configurations exist, the ground state
    is extensively degenerate. }
  \label{fig:AFI_LEL}
\end{figure}

We now apply our quantitative definition of frustration
(\Eq{eq:frustration}) to specific models. In \Fig{fig:AFI_LEL}, we
consider the antiferromagnetic Ising model, and compare $\epsilon^*$
with an usual local energy landscape representation of the
Hamiltonian. As we discussed in \Sec{sec:AFI}, this model only has
finite-range frustration: already at the scale of a three-sites
triangle, the model can be written in a frustration-free manner
(\Eq{eq:AFI_triangles}). Our approach consistently finds that this is
also true at the scale of the coordination shell: $f_7 = 0 $.

\begin{table*}[p]
  \centering
  \caption{ \label{tab:FLS} Frustration analysis of the Favoured Local
    Structures (FLS) model. Each line corresponds to a Hamiltonian
    where sites in the selected local structure -- the FLS -- are
    attributed an energy $-1$, while all other sites have zero
    energy. We consider all 13 distinct structures on the
    triangular lattice coordination shell (\Fig{fig:LSs}, the other 13
    being equivalent through spin inversion).  We identify  ground
    state structures and their energies $E_0$ using the systematic search
    algorithm described in Ref.~\cite{ronceray_favoured_2015}. The
    plots show two equivalent LEL representations of the Hamiltonian
    on the coordination shell cluster: its original definition as a
    delta-peaked LEL (blue dashes), and $\epsilon^*$, (one of) the LEL with
    highest minimal energy (orange line). We finally indicate the
    frustration defined in \Eq{eq:frustration} at the scale of three
    different clusters. This optimization is performed using the SciPy
    implementation of the Nelder-Mead
    algorithm~\cite{gao_implementing_2012}. }
  \setlength{\tabcolsep}{8pt}
  \begin{tabularx}{\textwidth}{cm{1.1cm}m{2.2cm}@{}c@{}c@{}m{4.5cm}ccc}
    \textbf{ID} & \multicolumn{1}{c}{\textbf{FLS}}  & \hspace{-.5em}\textbf{Ground state} & $E_0$ & \textbf{Cell size} & \qquad\textbf{Local energy landscape} & $f_7$ \includegraphics[scale=0.25]{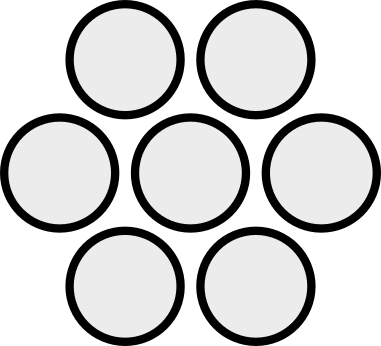} &  $f_{10}$  \includegraphics[scale=0.25]{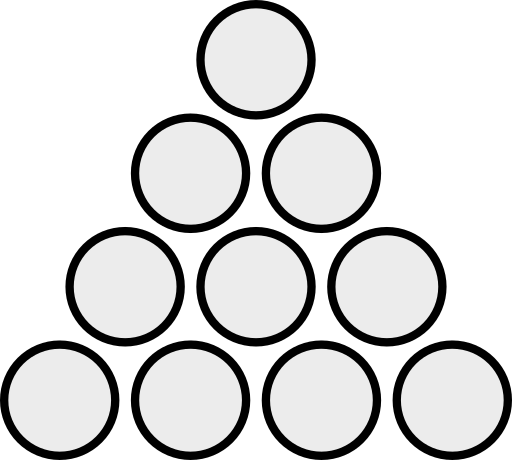} &  $f_{13}$  \includegraphics[scale=0.25]{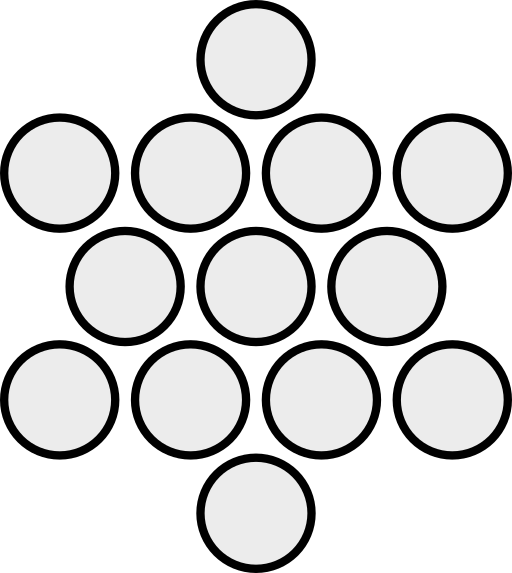} \\
    \midrule
    \input{FLS_data}
\end{tabularx}
\end{table*}

A more complex class of models is presented in \Tab{tab:FLS}: the
Favoured Local Structures (FLS) model. This model is defined directly
through its local energy landscape, which is delta-peaked to favor a
single structure (the FLS, with energy $-1$) while all others have
zero energy.  We have previously studied this model on
two-~\cite{ronceray_variety_2011} and
three-dimensional~\cite{ronceray_favoured_2015} lattices, revealing
that the geometry of the FLS controls a surprisingly rich
phenomenology, including complex crystalline ground
states~\cite{ronceray_favoured_2015}, liquid-liquid
transitions~\cite{ronceray_influence_2013,ronceray_multiple_2014}, and
slow dynamics~\cite{ronceray_multiple_2014,ronceray_favoured_2015}. In
\Tab{tab:FLS}, we summarize results for the variant of this model
where local structures are defined on the coordination
shell~\footnote{In
  Refs.~\cite{ronceray_variety_2011,ronceray_favoured_2015}, the local
  cluster considered are empty coordination shells, where the spin
  value at the central site is indifferent. This choice was made to
  limit the number of possible structures. However, in view of the
  results presented in this article, including the central spin is a
  more natural choice.}. Twelve of the thirteen distinct choices of
FLS (the exception being the trivial all-up case, labelled 0) result
in non-trivial ground states with $E_0 > -1$ (\emph{i.e.\@} including
non-FLS defects), and would thus be traditionally tagged as
frustrated. However, applying our formalism to each local energy
landscape, we find that in nine of these twelve systems, frustration
has a finite range and vanishes at the scale of the coordination shell
itself: $f_7=0$. While these systems appear frustrated, they are thus
completely equivalent to models with the same interaction range,
but for which there are multiple minima to the LEL, and a combination
of them can tile the lattice perfectly. The remaining three systems
(corresponding to structures 6, 9 and 11) have nonzero $f_7$, although
the numerical values for this frustration parameter (respectively
$1/12$, $2/117$ and $1/22$) are substantially smaller than the gap
between the FLS energy and the ground state energy in the initial
formulation of the problem (respectively $1/3$, $5/9$ and
$1/2$).  Thus even in the FLS model, a model
explicitly built to study geometrical frustration of local structures,
most systems that are apparently frustrated do not resist a closer
investigation: our gauge-invariant algorithm to quantify frustration
shows that frustration is an exception, rather than the norm.

\subsection{The range of frustration}
\label{sec:range}

Our quantitative definition of frustration (\Eq{eq:frustration})
depends on the size $z$ and geometry of the cluster on which we define
the LEL. This cluster must be at least as large as the range of
interactions of the Hamiltonian; it can however be
larger. Hierarchically increasing the cluster size by including more
sites, as in the sequence shown in \Tab{tab:clusters}, gives access to
more energy displacements, which are less local as they displace
energy over a longer range. The optimization in \Eq{eq:frustration}
thus occurs on a higher dimensional space when $z$ increases; as a
result, $f_z$ is a non-increasing function of $z$ when considering a
hierarchical family of clusters.

Our formalism thus distinguishes two classes of frustrated systems:
\begin{itemize}
\item \emph{systems with finite-range frustration} have a
  characteristic size $z^*$ such that $f_{z^*}=0$. This size
  corresponds to the scale at which the Hamiltonian can be written in
  an unfrustrated manner, \emph{i.e.\@} such that locally preferred
  structures can tile the whole space. At this scale, the locally
  preferred structures are typically degenerate, which can result in a
  degeneracy of the ground state of the system. Geometrical
  constraints are localized at scales $z \leq z^*$, and can be
  eliminated by an exact coarse-graining step.
\item \emph{systems with long-range frustration} have a nonzero $f_z$ at all
  scales: there is no way to write them in terms of finite-range
  unfrustrated LEL. Such systems would thus have truly non-local
  geometrical constraints. Stability of the ground state implies that the
  frustration function still decreases with scale, with an upper bound
  $f_z < A / z^{1/d}$ where $d$ is the dimension of space~\footnote{One
    type of energy displacement~$\delta$ is to average the energy over
    clusters in a ball of radius~$R$ (with $z\sim R^d$ sites).  The
    resulting LEL $\epsilon+\delta$ on the ball cannot admit a
    configuration with energy less than $E_0-O(1/R)$, otherwise
    replacing the region of the ground state by that configuration
    would lower the energy $E_0 R^d$ more than the energy cost
    $A R^{d-1}$ of the boundary.}.
\end{itemize}

On the triangular lattice, we have seen that the antiferromagnetic
Ising model has finite-range frustration with $z^* = 3$
(\Eq{eq:AFI_triangles}), while nine of the twelve frustrated FLS
models have $z^* = 7$ (\Tab{tab:FLS}). Interestingly, we find that the
remaining three choices of FLS have finite-range frustration too, with
$z^* =10$ for FLSs $6$ and $11$, and $z^*=13$ for FLS $9$. Note that these
structures correspond to most of those for which the crystalline
ground state has the largest elementary cell (respectively $9$, $27$
and $12$). Thus, none of the systems defined by a FLS on the triangular
lattice coordination shell have long-range frustration. Extending this
study to the case of a chiral Hamiltonian (\emph{i.e.\@} favoring only
structure 12, but not its enantiomer) and to the case of structures
defined on the empty shell (with a cluster including the six neighbors
of a site, but not the site itself, as studied in
Refs.~\cite{ronceray_variety_2011,ronceray_influence_2013,ronceray_multiple_2014})
does not change this conclusion: all 2D binary spin systems studied by
the authors have $z^* \leq 13$, and thus have finite-range frustration
only. At the time of this writing, it remains unclear whether
there exists discrete spin systems with long-range frustration.

A practical constraint to the investigation of more complex structures
(\emph{e.g.\@} with more than two spin values, or on three-dimensional
lattices) is that the number $n$ of structures grows exponentially
with $z$. The energy displacements are obtained as the null space of
the $n\times n$ matrix $\mathbf{C}$; we conjecture that their number
grows exponentially too (\Tab{tab:clusters}). This puts sharp
constraints on the size at which it is possible to study frustration
with our method; in particular, any type of scaling analysis is
impossible for now. This might not be hopeless, though: in this
article, our search through the energy displacement space is
blind. Identifying in advance what energy displacement will matter
could allow to estimate $f_z$ without having to perform the
high-dimensional optimization. We leave this possibility open for
future work.

\subsection{Provability of ground states}
\label{sec:GS}

To finish on a brighter note, we present a practical application of
our framework in the identification of ground state
energies. Computing the ground state energy of a many-body spin
Hamiltonian such as the FLS models (\Tab{tab:FLS}) is a challenging
problem, even if their interactions are short-ranged. In practice, we
have found that constructive, enumerative techniques permit to
investigate all possible crystalline structures up to a given cell
size, using an adaptation of the algorithm developed by Hart and
Forcade~\cite{hart_algorithm_2008,ronceray_favoured_2015}. This
algorithm typically provides a ``good candidate'' for the ground state
structure.  However, it is difficult to know for sure that this
candidate is, indeed, the ground state: how to be sure that no
structure with a larger, more complex unit cell and a slightly lower
energy exists?

Our framework provides lower bounds to this ground state energy: the
LPS energy of any LEL representation of the Hamiltonian
(\Eq{eq:E0_bound}). In particular, for a given cluster size $z$ on
which we define the LEL, the most restrictive bound is
\begin{equation}
  \label{eq:Estar}
  E^*_z(\epsilon) =  \max_{\delta\in\Delta}\left[\min_{\stra=1\dots n}(\epsilon_\stra+\delta_\stra) \right]
\end{equation}
\emph{i.e.\@} the maximal LPS energy in \Eq{eq:frustration}. When this
energy $E^*_z(\epsilon)$ coincides with the energy of a crystalline
state that could be constructed with, \emph{e.g.\@}, our enumerative
algorithm, it means that the system has finite-range
frustration. Furthermore, it provides a rigorous proof that this state
is, indeed, the ground state of the system. As all FLS systems
presented in \Tab{tab:FLS} have finite-range frustration, we thus have
proven that the crystalline structures depicted in this table are,
indeed, ground state configurations. This also applies to the variant
of the model studied in
Refs.~\cite{ronceray_variety_2011,ronceray_influence_2013,ronceray_multiple_2014}.

Interestingly, this method of ``proving ground state energies'' would
not work for systems with long-range frustration (if such systems
exist). This could mean that these systems effectively belong to a
different class of complexity for the provability of their ground
states.

\section{Discussion}
\label{sec:discussion}

In this article, we have examined the notion of geometrical
frustration in the context of lattice spin models with short-range
interactions and translation invariance. This notion is understood
here as the impossibility for the locally preferred order to tile
space.  To sharpen this idea of locally preferred order, we introduce
the framework of \emph{local energy landscapes}, which associates an
energy to each spin depending on its local spin environment --
\emph{i.e.\@} its local structure. There is, however, an ambiguity in
this choice: for a given Hamiltonian, we have seen that there are
typically many equivalent ways to define a local energy landscape,
related by unphysical gauge changes that we term \emph{energy
  displacements}. We have shown how to characterize the gauge group,
and construct it in practical cases, using a high-temperature
expansion of the entropy. This allows us to define a gauge invariant
measure for frustration, which depends only on the Hamiltonian and the
scale considered. The scale-dependence of this frustration function
defines two classes of frustrated systems. When frustration vanishes
above a certain scale, we say that the system has \emph{finite-range
  frustration}: it can be eliminated by a local ``blurring'' of the
local energy. All systems studied in this article fall in this class;
in such cases, our framework provides a rigorous proof that our
estimate of the minimum energy is indeed the ground state of these
systems. We speculate that a second class of systems, that we term
\emph{long-range frustrated}, exists. For such systems, the
geometrical constraints in their organization are non-local, which
might lead to interesting physical properties; however, an example of
spin system with long-range frustration remains to be discovered.

The key difficulty in this assessment of geometrical frustration is
that it requires a notion of \emph{local energy}, which is typically
ambiguous: only the global Hamiltonian has a true physical
meaning. The gauge of energy displacements, that we have characterized
here in the case of spin systems, reflects this ambiguity: two local
energy landscapes related by an energy displacement are virtually
indistinguishable. This has practical consequences: any attempt to
infer the LEL from experimental measurements of the statistics of
structures would be unable to resolve such difference, and would thus
yield ambiguous results. Our framework resolves this ambiguity.  Other
approaches attempting to attribute energies to local structures, for
instance in particle systems in the study of icosahedral
structures~\cite{frank_supercooling_1952,tarjus_frustration-based_2005,taffs_role_2016}
or other clusters~\cite{malins_identification_2013}, might be subject
to such ambiguity too. Our framework could be adapted to such systems,
and provide a route towards a quantitative measure of
frustration. This is, we argue, a necessary step towards connecting
geometrical frustration to its alleged consequences, such as the
extensive degeneracy of ground states or slow dynamics in the
supercooled liquid.

Finally, we note that we have only considered bulk systems here, for
which there is no need to specify boundary conditions. This is the
relevant case for the thermodynamic properties of liquids, crystals
and glasses.  However, there has been recently an emergent interest in
the physical properties of geometrically frustrated systems with free
boundaries, such as assembling proteins or filaments in a dilute
solution~\cite{grason_perspective:_2016,lenz_geometrical_2017}. In
this ``geometrically frustrated assembly'' paradigm, frustration in
the bulk competes with surface tension at the free surface. In order
to apply our framework to these systems, one would thus need to
consider the influence of energy displacements on the surface tension.

\vspace{5mm}

\begin{acknowledgments}
The authors thank Peter Harrowell, Gilles
Tarjus, Anna Frishman, Ricard Alert-Zenon, Martin Lenz
and Francesco Turci for useful conversations and stimulating
comments. PR is supported by a Princeton Center for
Theoretical Science fellowship.
\end{acknowledgments}

\bibliography{Frustration}

\end{document}

%% file: clusters_table.tex
\begin{table*}[ht]
  \centering
  \setlength{\tabcolsep}{5pt}
  \begin{tabularx}{\textwidth}{r@{}c@{\hspace{-1em}}c@{\hspace{-1em}}c@{}cccc}
    \hline
    \raisebox{20pt}{\textbf{Cluster}}  &
    \includegraphics[scale=0.4]{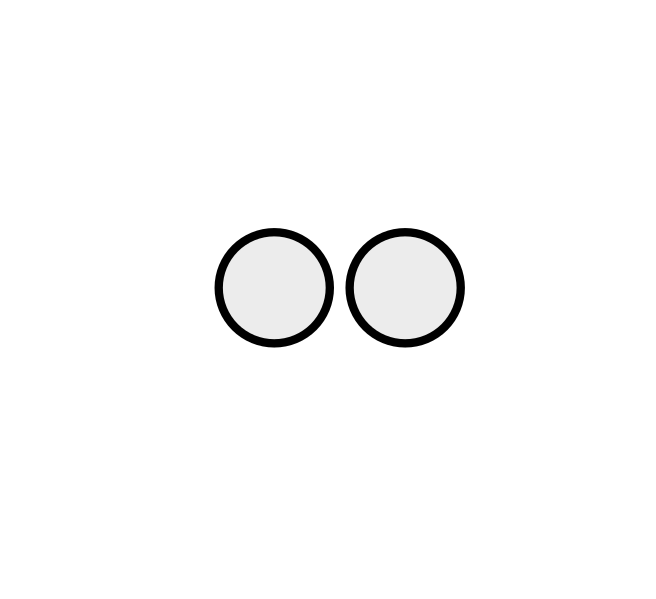} &
    \includegraphics[scale=0.4]{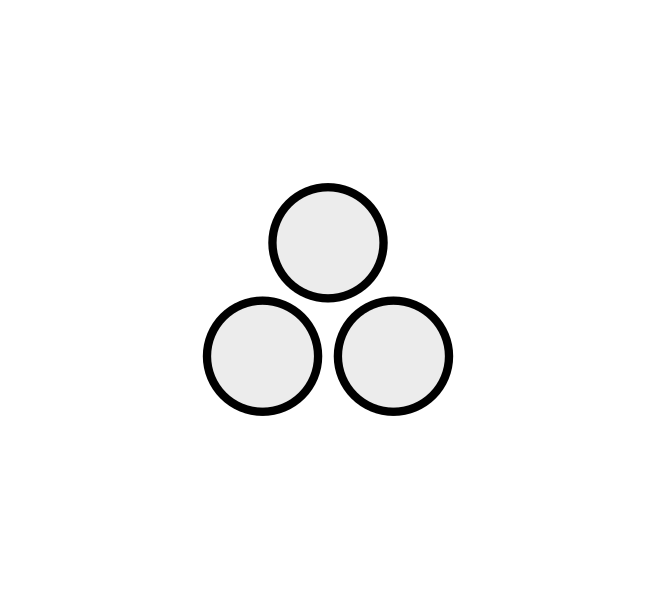} &
    \includegraphics[scale=0.4]{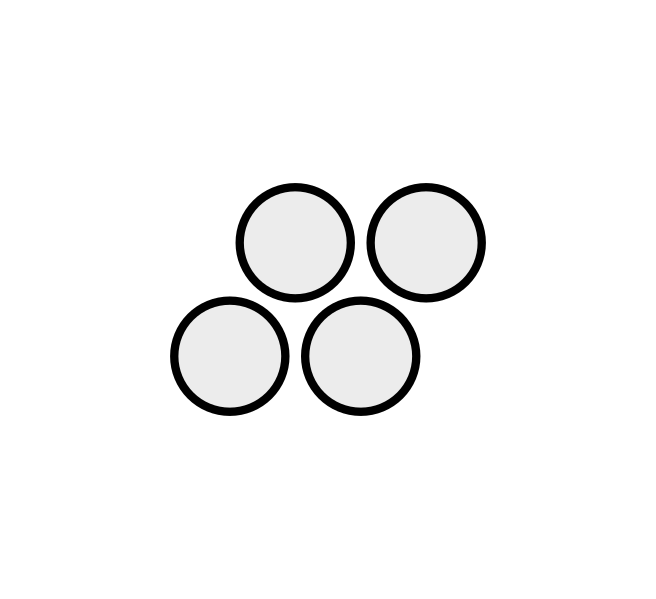} &
    \includegraphics[scale=0.4]{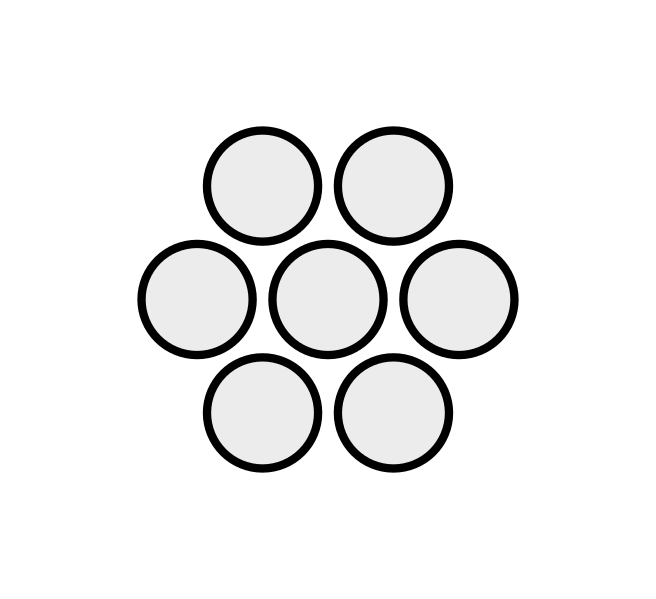} &
    \includegraphics[scale=0.4]{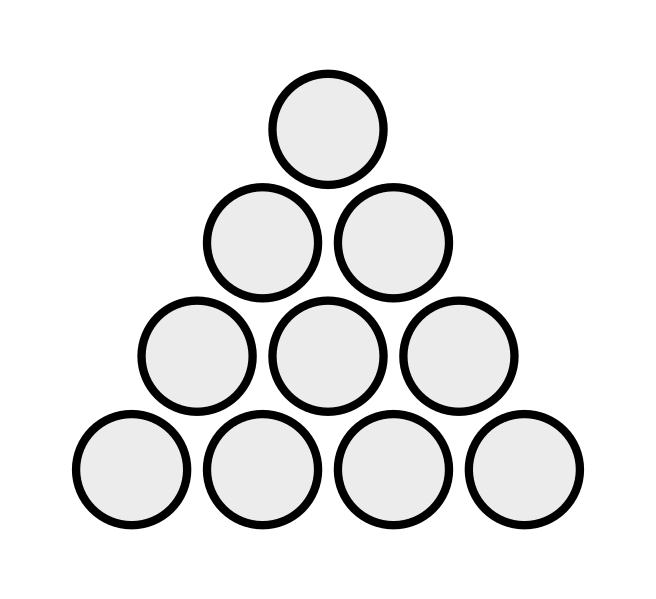} &
    \includegraphics[scale=0.4]{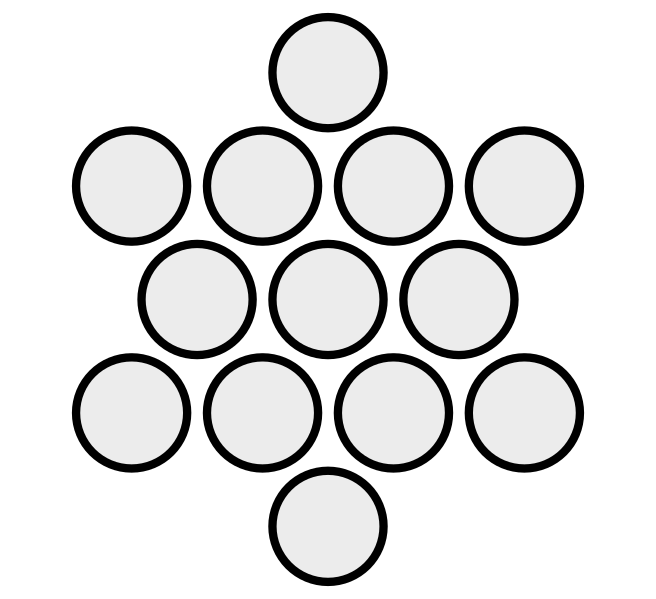} &
    \includegraphics[scale=0.4]{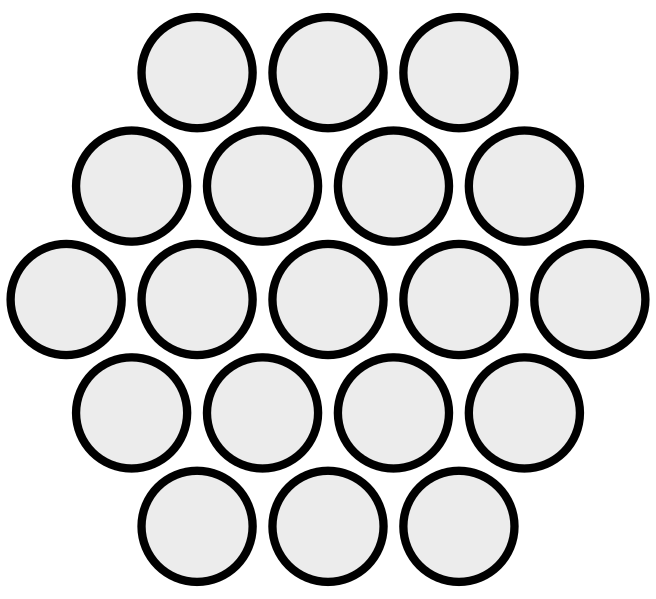} \\

    \textbf{Size} $z$ &
    2 &
    3 &
    4 &
    7 &
    10 &
    13 &
    19 \\
    
    \textbf{Number of LSs} $n$ &
    3 &
    4 &
    9 (10) &
    26 (28) &
    208 (352) &
    828 (1,400) &
    45,336 (87,600) \\
    
    \textbf{Energy displacements} $\dim\Delta$ &
    0 &
    0 &
    2 &
    3 (3) &
    20 (37) &
    59 (103) &
    3504 (6753) \\
    \hline
\end{tabularx}
\caption{ \label{tab:clusters} High-symmetry local clusters on the
  triangular lattice, and the properties of corresponding local energy
  landscapes. Numbers in brackets correspond to chiral cases,
  \emph{i.e.} considering enantiomeric structures as being distinct.
}
\end{table*}

%% file: FLS_data.tex
0 & \includegraphics[width=0.06\textwidth]{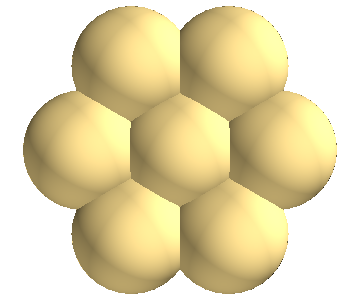} & \includegraphics[width=0.1\textwidth]{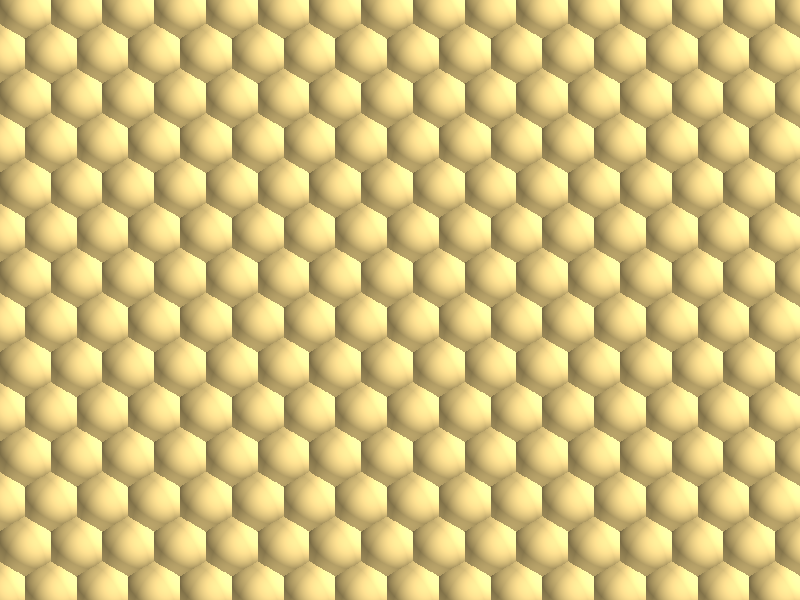} & $-1$ & $1$ & \includegraphics[width=0.25\textwidth]{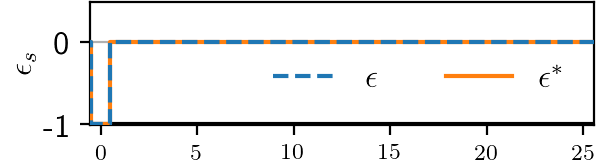}& $0$ & $0$ & $0$ \\1 & \includegraphics[width=0.06\textwidth]{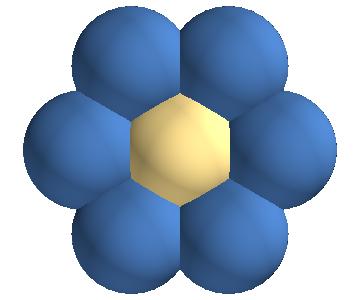} & \includegraphics[width=0.1\textwidth]{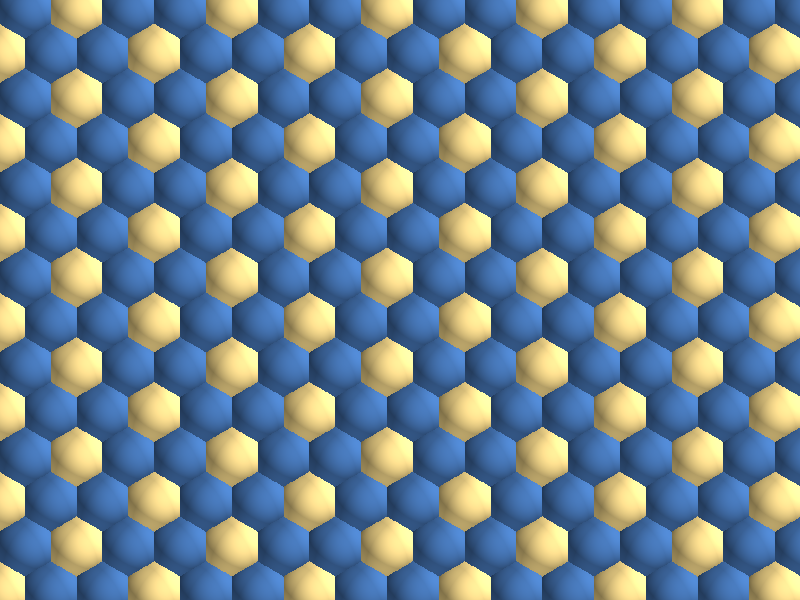} & $-0.3333$ & $3$ & \includegraphics[width=0.25\textwidth]{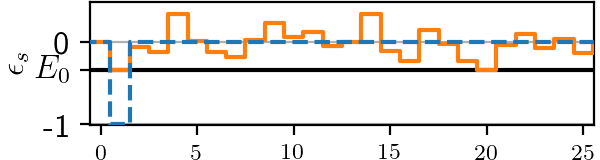}& $0$ & $0$ & $0$ \\2 & \includegraphics[width=0.06\textwidth]{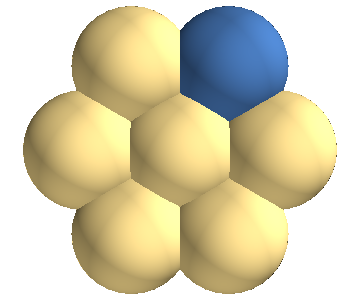} & \includegraphics[width=0.1\textwidth]{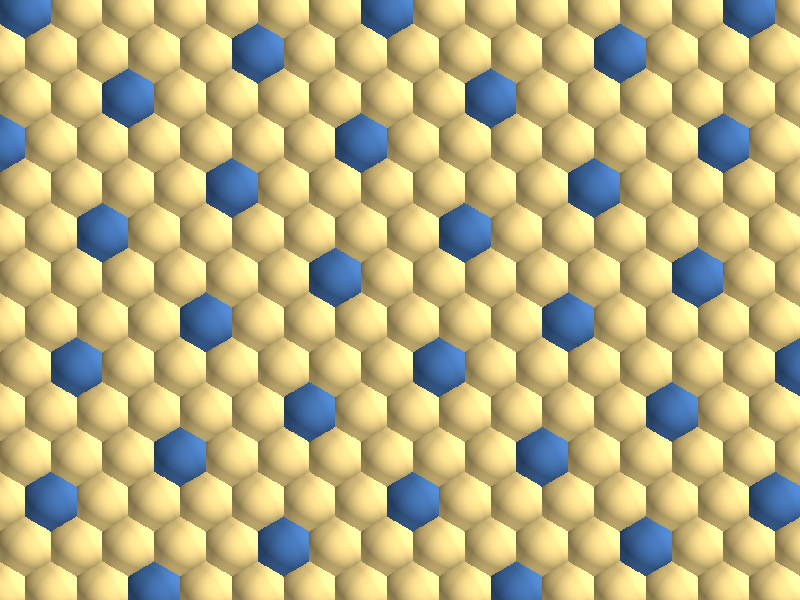} & $-0.8571$ & $7$ & \includegraphics[width=0.25\textwidth]{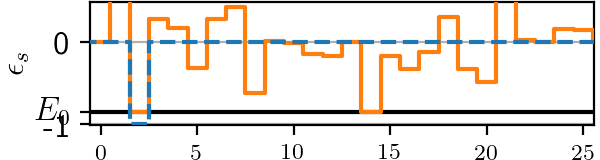}& $0$ & $0$ & $0$ \\3 & \includegraphics[width=0.06\textwidth]{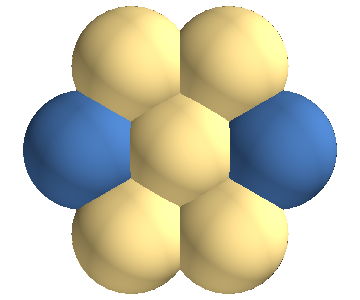} & \includegraphics[width=0.1\textwidth]{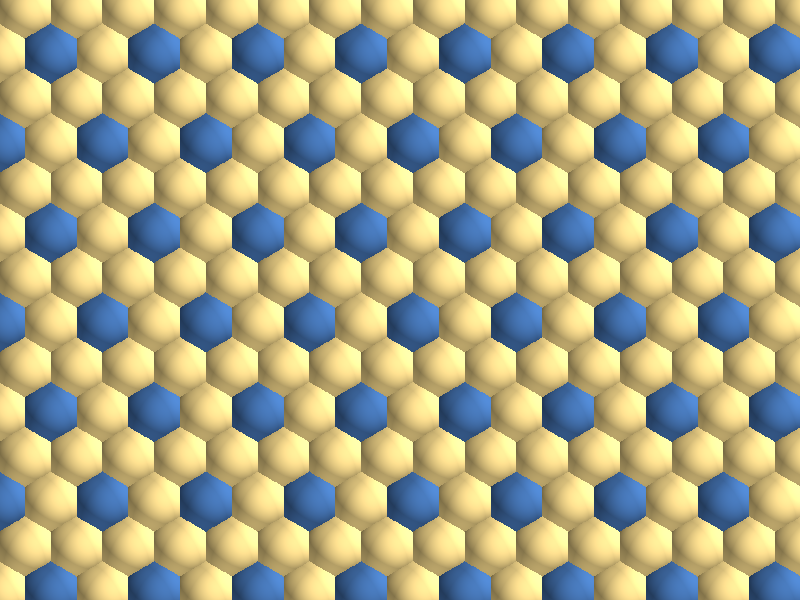} & $-0.75$ & $4$ & \includegraphics[width=0.25\textwidth]{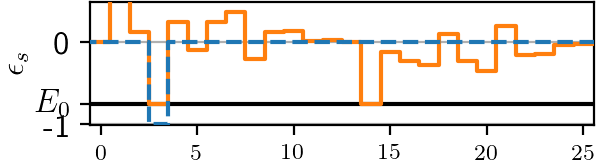}& $0$ & $0$ & $0$ \\4 & \includegraphics[width=0.06\textwidth]{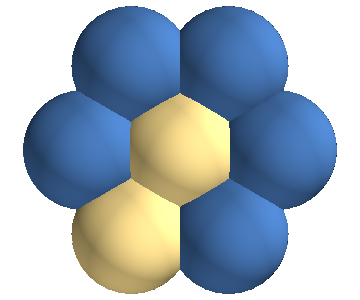} & \includegraphics[width=0.1\textwidth]{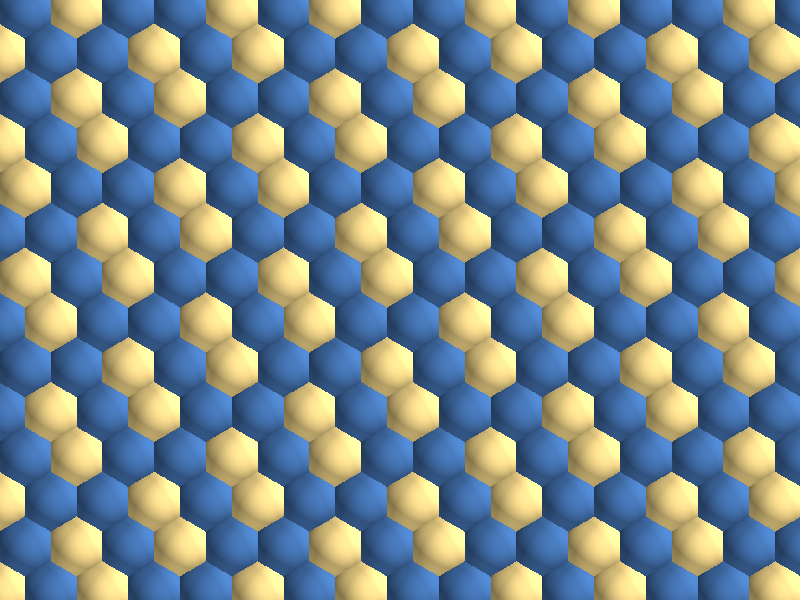} & $-0.4$ & $5$ & \includegraphics[width=0.25\textwidth]{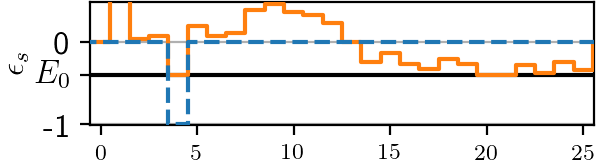}& $0$ & $0$ & $0$ \\5 & \includegraphics[width=0.06\textwidth]{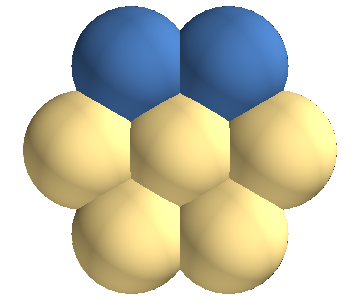} & \includegraphics[width=0.1\textwidth]{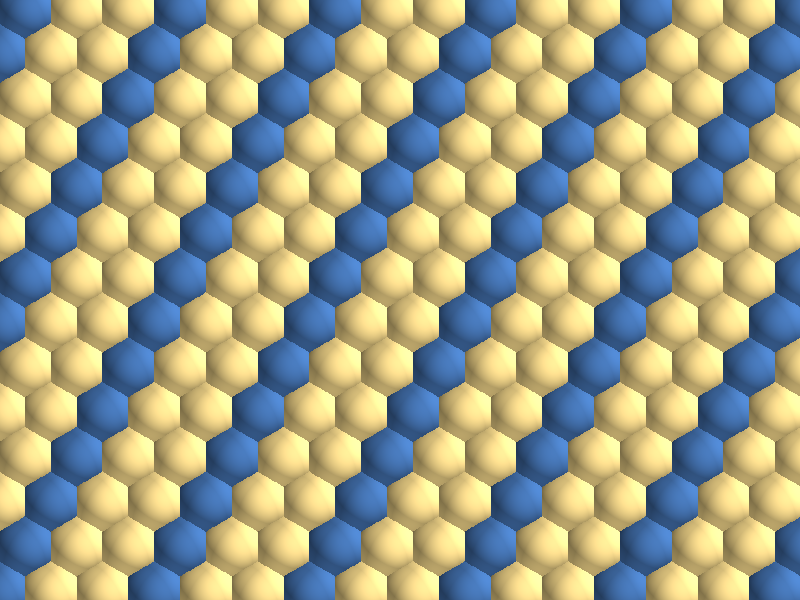} & $-0.6667$ & $3$ & \includegraphics[width=0.25\textwidth]{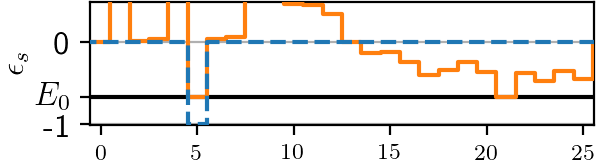}& $0$ & $0$ & $0$ \\6 & \includegraphics[width=0.06\textwidth]{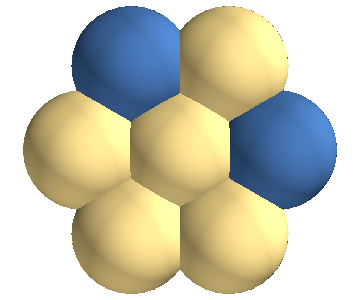} & \includegraphics[width=0.1\textwidth]{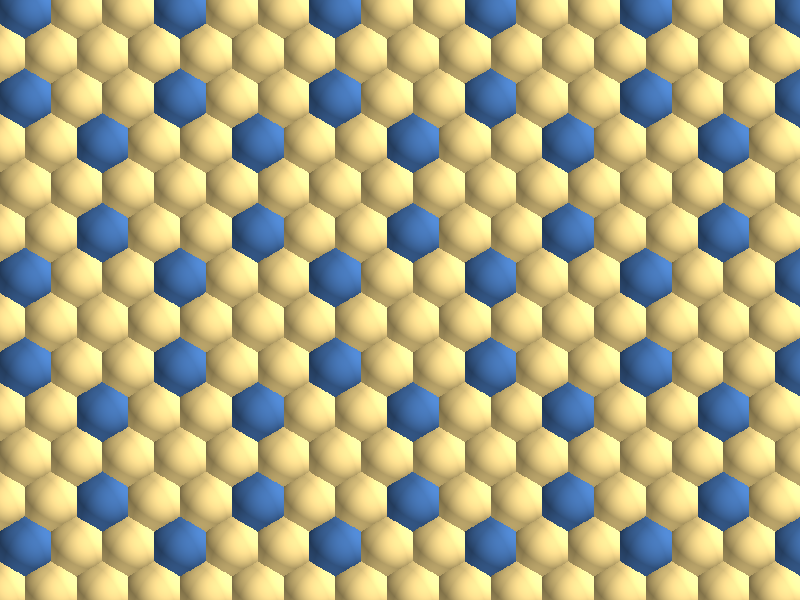} & $-0.6667$ & $9$ & \includegraphics[width=0.25\textwidth]{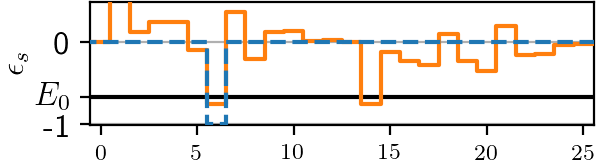}& $0.0833$ & $0$ & $0$ \\7 & \includegraphics[width=0.06\textwidth]{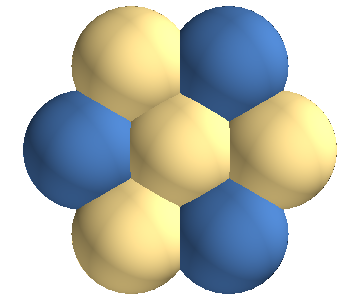} & \includegraphics[width=0.1\textwidth]{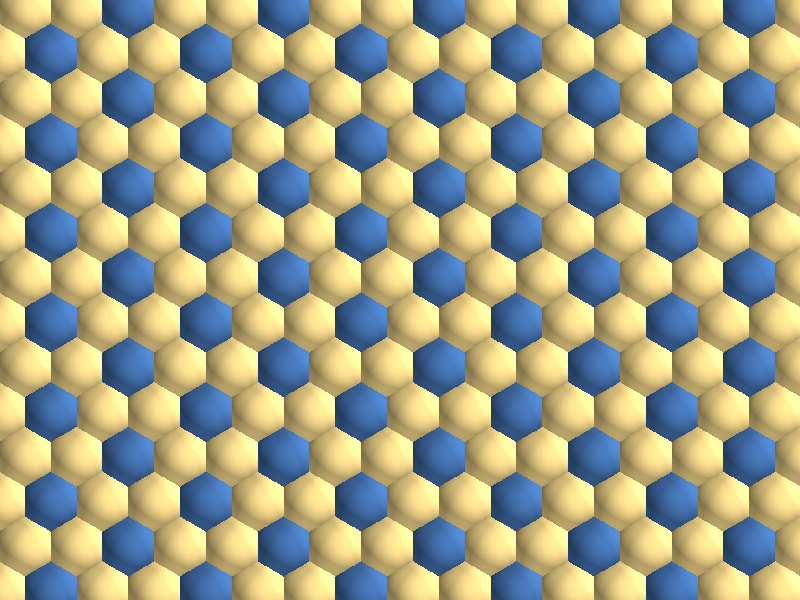} & $-0.6667$ & $3$ & \includegraphics[width=0.25\textwidth]{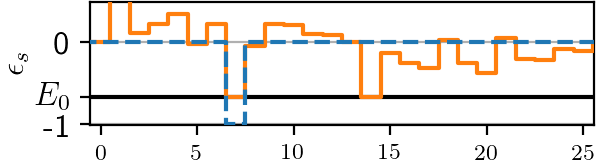}& $0$ & $0$ & $0$ \\8 & \includegraphics[width=0.06\textwidth]{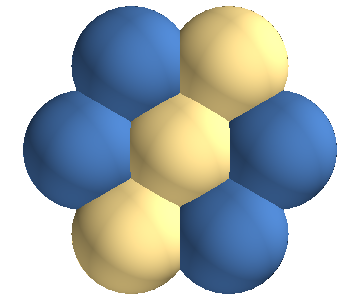} & \includegraphics[width=0.1\textwidth]{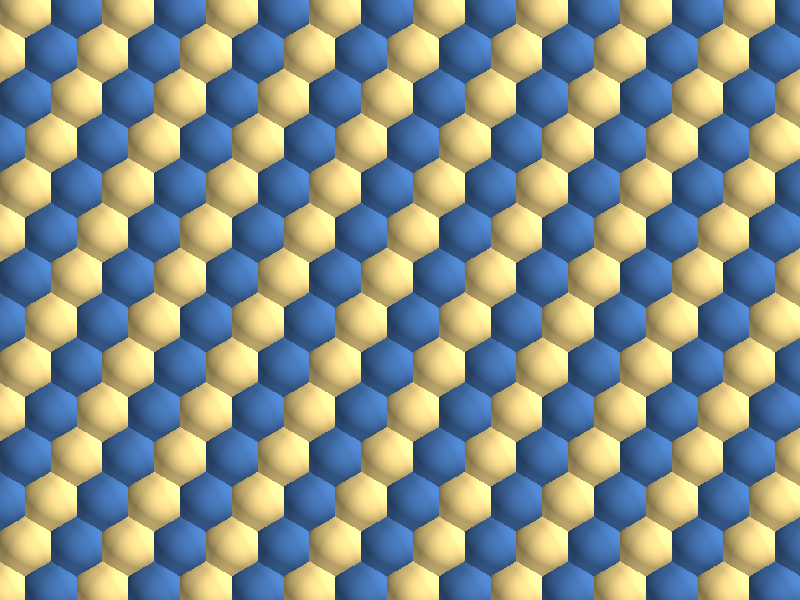} & $-0.5$ & $2$ & \includegraphics[width=0.25\textwidth]{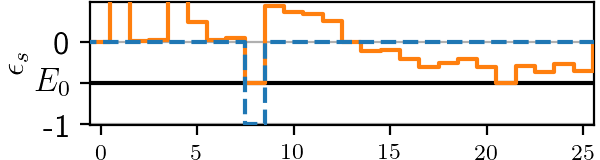}& $0$ & $0$ & $0$ \\9 & \includegraphics[width=0.06\textwidth]{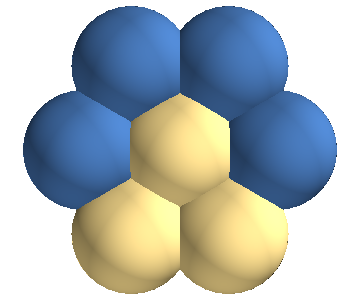} & \includegraphics[width=0.1\textwidth]{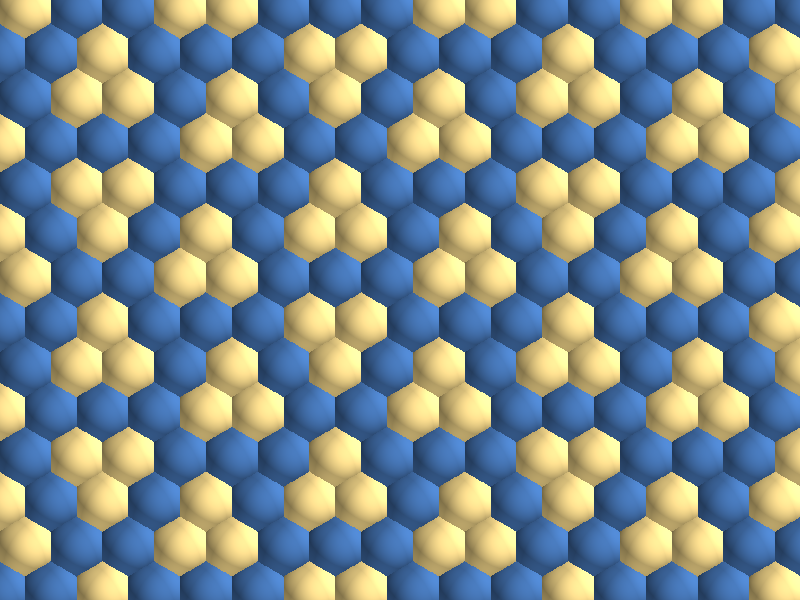} & $-0.4444$ & $27$ & \includegraphics[width=0.25\textwidth]{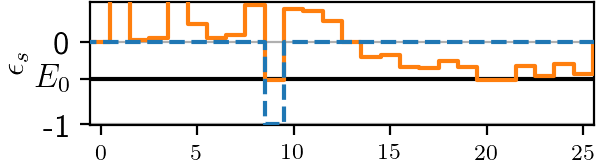}& $0.0171$ & $0.0171$ & $0$ \\10 & \includegraphics[width=0.06\textwidth]{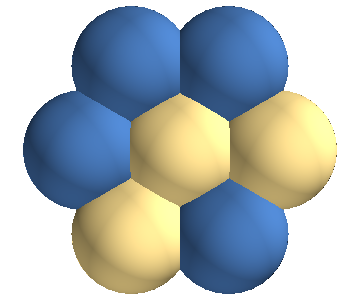} & \includegraphics[width=0.1\textwidth]{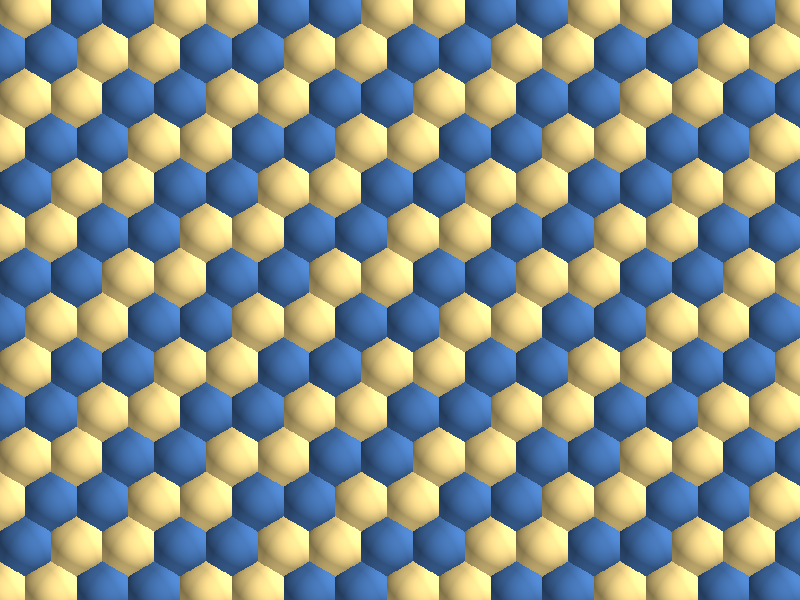} & $-0.5$ & $4$ & \includegraphics[width=0.25\textwidth]{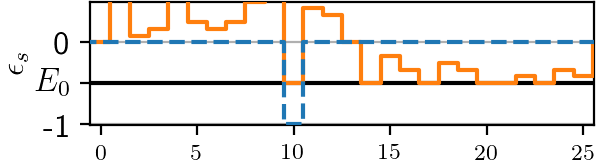}& $0$ & $0$ & $0$ \\11 & \includegraphics[width=0.06\textwidth]{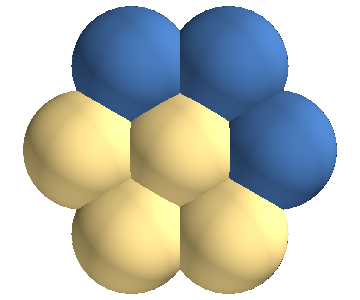} & \includegraphics[width=0.1\textwidth]{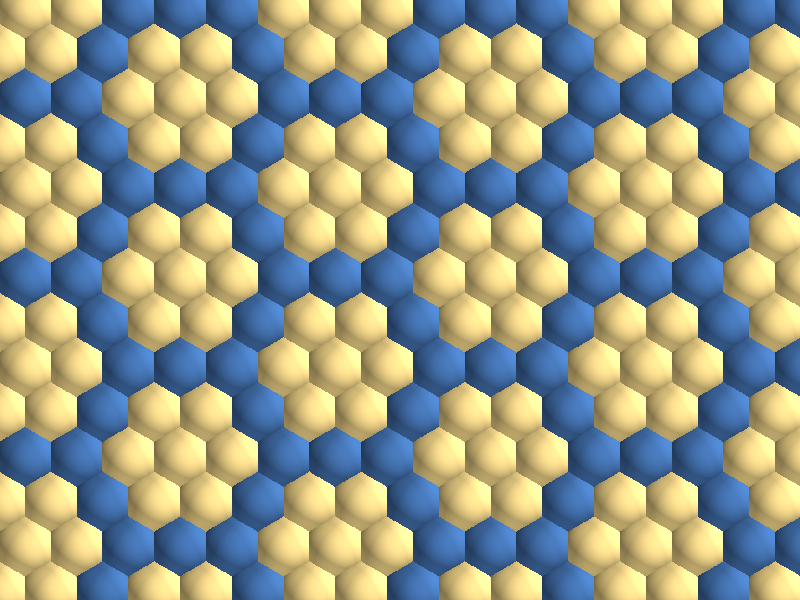} & $-0.5$ & $12$ & \includegraphics[width=0.25\textwidth]{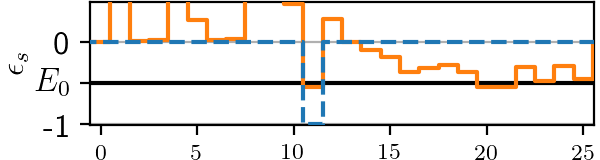}& $0.0455$ & $0$ & $0$ \\12 & \includegraphics[width=0.06\textwidth]{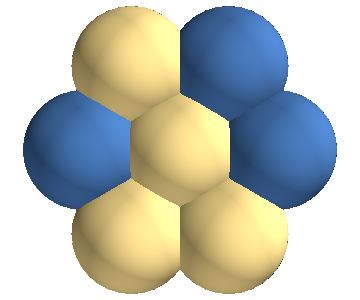} & \includegraphics[width=0.1\textwidth]{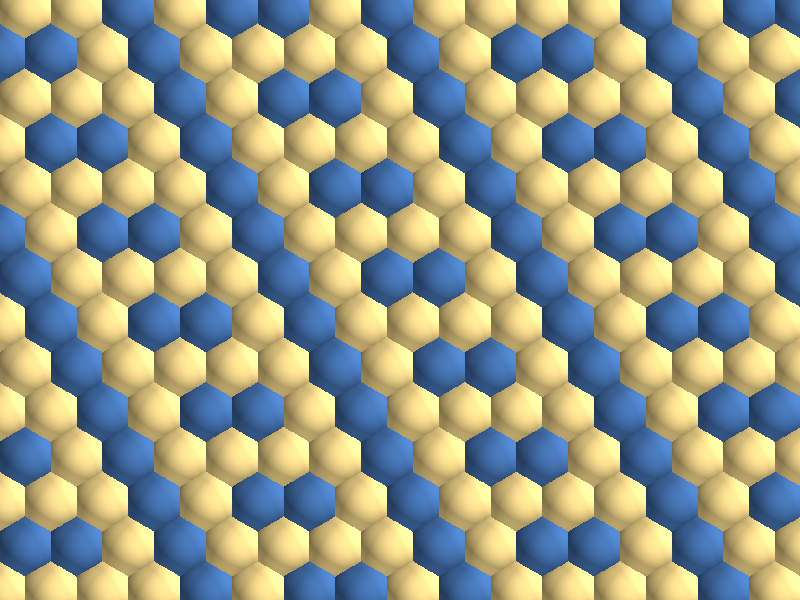} & $-0.6$ & $10$ & \includegraphics[width=0.25\textwidth]{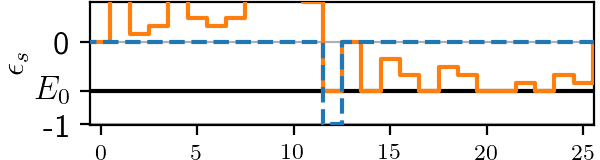}& $0$ & $0$ & $0$ \\